\documentclass[sigconf,nonacm]{acmart}

\renewcommand\footnotetextcopyrightpermission[1]{}

\usepackage{subcaption}

\usepackage{amsmath,amsfonts,bm}

\def\eqref#1{equation~\ref{#1}}

\def\1{\bm{1}}

\DeclareMathAlphabet{\mathsfit}{\encodingdefault}{\sfdefault}{m}{sl}
\SetMathAlphabet{\mathsfit}{bold}{\encodingdefault}{\sfdefault}{bx}{n}

\usepackage{pifont}
\usepackage{booktabs}
\usepackage{graphicx}
\usepackage{listings}
\usepackage{tcolorbox}
\usepackage{multirow}
\usepackage{array}
\usepackage{xspace}
\usepackage{tabularx}
\usepackage{makecell}
\usepackage{booktabs}
\usepackage{multirow}
\usepackage{placeins}

\newcolumntype{Y}{>{\centering\arraybackslash}X}

\newcommand{\find}[1]{
\begin{tcolorbox}[leftrule=1mm,toprule=0mm,bottomrule=0mm,left=1pt,right=2pt,top=2pt,bottom=2pt]
\em #1
\end{tcolorbox}
}

\newcommand{\bench}{\textsc{CrossPloit}\xspace}
\newcommand{\vsw}{\textsc{VulScope-Web}\xspace}

\usepackage{listings}
\lstdefinestyle{trajectory}{
  language=Python,
  basicstyle=\ttfamily\scriptsize,
  columns=fullflexible,
  breaklines=true,
  breakatwhitespace=false,
  keepspaces=true,
  showstringspaces=false,
  showspaces=false,
  showtabs=false,
  tabsize=2,
  frame=single,
  numbers=left,
  numberstyle=\tiny\color{gray},
  numbersep=5pt,
  captionpos=b,
  aboveskip=0.5\baselineskip,
  belowskip=0.5\baselineskip,
  keywordstyle=\color{blue},
  commentstyle=\color{green!40!black},
  stringstyle=\color{orange!80!black},
  escapeinside={(*@}{@*)},
  moredelim=**[is][\color{red}\ttfamily]{@}{@},
  moredelim=**[is][\color{red}\ttfamily]{<<}{>>},
  moredelim=**[is][\color{blue}\ttfamily]{[[}{]]}
}
\definecolor{dimgreen}{RGB}{34, 139, 34} 
\definecolor{darkblue}{RGB}{0,51,153}

\begin{document}

\title{\bench: A Benchmark for Cross-Version PoC Migration in Python Web Applications}

\author{Hung Nguyen-Huu}
\affiliation{%
  \institution{Singapore Management University}
  \city{Singapore}
  \country{Singapore}
}

\author{Nguyen Anh Tuan}
\affiliation{%
  \institution{Hanoi University of Science and Technology}
  \city{Hanoi}
  \country{Vietnam}
}

\author{Ting Zhang}
\affiliation{%
  \institution{Monash University}
  \city{Melbourne}
  \country{Australia}
}
\authornote{Ting Zhang is the corresponding author.}

\author{Yue Liu}
\affiliation{%
  \institution{Singapore Management University}
  \city{Singapore}
  \country{Singapore}
}

\author{Chengran Yang}
\affiliation{%
  \institution{Singapore Management University}
  \city{Singapore}
  \country{Singapore}
}

\author{Lwin Khin Shar}
\affiliation{%
  \institution{Singapore Management University}
  \city{Singapore}
  \country{Singapore}
}

\author{Eng Lieh Ouh}
\affiliation{%
  \institution{Singapore Management University}
  \city{Singapore}
  \country{Singapore}
}

\author{David Lo}
\affiliation{%
  \institution{Singapore Management University}
  \city{Singapore}
  \country{Singapore}
}

\renewcommand{\shortauthors}{Nguyen-Huu et al.}

\begin{abstract}
Affected-version ranges in vulnerability databases decide which deployments scanners, patching pipelines, and incident responders treat as vulnerable, yet these ranges are often incomplete, so vulnerable releases are silently treated as safe. Auditing them requires executable evidence: given a proof-of-concept (PoC) exploit for a listed version, show that the same vulnerability can be triggered on an earlier, unlisted version. This cross-version PoC migration combines source-code analysis, environment repair, and execution-based verification. We introduce \bench, a benchmark of 127 migration instances across 45 CVEs in the Python web ecosystem. Each instance includes a version-pinned container, a reference PoC, and a vulnerability-specific oracle; two validators check that the observed effect arises from the disclosed vulnerability. Benchmark construction confirms 149 vulnerable prior versions absent from the affected-version lists of four databases, 127 of which require migration. Across 18 agent configurations (nine models and two harnesses), the best configuration resolves 58.3\% of instances, compared with 8.7\% for an adapted trace-guided baseline. Attack-vector shifts are particularly difficult: no configuration resolves more than 37.9\% of these instances. Higher resolve rates are associated with a smaller share of missed-API failures and a larger share of incomplete executions among the remaining failures. In one configuration, withholding the reference PoC and its environment specification reduces resolve rate from 50.4\% to 15.0\%, highlighting the value of the supplied reference package. \bench provides an execution-based measurement of under-reported exposure and of how far its enumeration can be automated.
\end{abstract}



\maketitle

\section{Introduction}
\label{sec:intro}
Modern vulnerability management rests on an assumption: a software version outside a CVE's listed affected-version range is not vulnerable.
When a vulnerability is disclosed, databases such as the National Vulnerability Database (NVD)~\cite{NVDHome30:online}, OSV~\cite{osv}, GHSA~\cite{ghsa}, and Snyk~\cite{snyk} record which versions are affected, and the defensive stack built on top of them inherits that boundary as ground truth.
Software composition analysis (SCA) scanners flag a deployment only when its version falls inside the range; patching pipelines schedule upgrades against the same boundary; and incident responders triage exposure by asking whether a running version is listed.
A version just outside the range is, in effect, declared safe by every tool a defender trusts.

This boundary is frequently wrong, and, more importantly for a defender, it is wrong \emph{invisibly}.
One line of prior work shows that affected-version ranges are routinely incomplete, but it \emph{infers} vulnerability statically from commit history or code similarity~\cite{cheng2026vercation,bao2022v,woo2021v0finder}: such analysis can flag a version as suspicious, yet it cannot confirm that the version is exploitable, and an unconfirmed signal is easy to discount.
A separate line produces executable evidence by migrating an exploit across versions, but it targets C/C++ binaries or Java libraries and relies on a crash or a fixed test assertion as its success signal~\cite{vulscope,jiang2023aem,chen2025diffploit}; neither the oracle nor the tooling transfers to the Python web setting.
The result is a blind spot the defensive stack cannot see by construction: an exploitable but unlisted version is reported clean by the scanner, skipped by the patching campaign, and dismissed during incident response, with no contradicting evidence anywhere in the pipeline.
The same blind spot is an adversary's opening: an attacker who holds a working exploit for a listed version has a direct incentive to show that an earlier, unlisted version falls to the same flaw, extending a known n-day onto exactly the deployments defenders believe are out of scope.
How much of this exposure exists in a real ecosystem, and how cheaply it can be enumerated, has not been measured with executable proof.

Confirming the exposure requires evidence rather than inference: given a proof-of-concept (PoC) exploit for a listed affected version, demonstrate that the same vulnerability can be triggered on an earlier version that is not listed.
We call this task \emph{cross-version PoC migration}: preserving the underlying vulnerability while accommodating the differences between versions.
A successful migration turns an unconfirmed gap into a concrete, reproducible artifact, namely a running exploit against a version the databases call safe.
Producing this evidence is hard for three reasons, and each is also why the exposure has gone unmeasured.
First, \textbf{reference PoCs are often unavailable or non-reproducible}: CVE descriptions do not yield runnable exploits, and public PoCs are sparse, so confirming any exposure first requires reconstructing the vulnerable application, pinning dependencies, and defining an observable success condition.
Second, \textbf{migration is not a replay}: across versions, APIs are renamed, request paths change, validators appear or disappear, and configuration defaults shift; in some cases the original entry point no longer reaches the vulnerable logic even though the flaw remains reachable through a different code path.
For instance, the denial of service of CVE-2024-45230~\cite{CVE-2024-45230} reaches the unlisted Django~4.0 only after the exploit abandons the reference PoC's vector for a different branch of the same routine (Section~\ref{app:motivation-example}).
Third, \textbf{web vulnerabilities lack a universal success oracle}: a memory-corruption bug announces itself with a crash, but a database disclosure, an unauthorized file read, and a denial of service manifest differently, so deciding whether a presumed-safe version is genuinely compromised demands a per-vulnerability oracle plus a check that rules out incidental errors, misconfigurations, and modifications that introduce the vulnerability into the target.

We address these challenges with \bench, a benchmark that doubles as an instrument for measuring under-reported exposure in the Python web ecosystem, covering Django, Flask, FastAPI, Web2py, and packages built on them.
From 217 candidate CVEs, we first reconstruct 97 validated reference PoCs on listed affected versions, each packaged in a containerized environment with a standardized PoC specification (addressing the first challenge).
Two annotators then independently attempt to migrate each reference PoC to prior unlisted versions, using a version window and time budget calibrated in a pilot study, before reconciling a canonical result and labeling the adaptation it required (addressing the second challenge).
Every accepted instance carries a deterministic, vulnerability-specific oracle, and two validators confirm not merely that the oracle fires but that it fires because the disclosed vulnerability is triggered on the target version (addressing the third challenge).
The resulting benchmark contains 127 migration instances across 45 CVEs, spanning API changes, payload tuning, attack-vector shifts, dependency repair, and configuration changes.

We organize the study around three research questions that follow the defender's view of the blind spot: how large it is, how far its enumeration can be automated, and where automation breaks down.
Benchmark construction itself answers the first question: it confirms \textbf{149 vulnerable prior versions absent from all four databases' affected-version lists}, that is, 149 deployments-worth of exposure that an SCA scanner, patching pipeline, or triage built on those lists would declare safe.
This exposure is mostly invisible to replay: only 22 of the 149 versions fall to the unmodified reference PoC, while 127 require migration, so an audit that merely re-runs published PoCs would recover about one-sixth of the exposure in our setting.
To answer the second question, we evaluate nine models in each of two agent harnesses, OpenHands and SWE-agent, together with an adaptation of the trace-guided tool VulScope~\cite{vulscope}.
The adapted baseline, which is restricted to input mutation, resolves only 11 of 127 instances (8.7\%).
LLM-based agents, which can also modify PoC code and environments, do better but remain far from complete: the strongest configuration, OpenHands with Opus-4.6, resolves 74 instances (58.3\%), and no configuration resolves more than 37.9\% of the instances that require an attack-vector shift.
For the third question, failure analysis shows that higher resolve rates are associated with a smaller share of missed-API failures and a larger share of incomplete executions among the remaining failures; these associations characterize the evaluated configurations rather than a limit on future models.
Under-reported version exposure is therefore a measurable, partially automatable security gap whose barrier to enumeration is falling but not yet low, for defenders and attackers alike.

In summary, this paper makes the following contributions:
\begin{itemize}
  \item \textbf{A measurement of a defender blind spot.} With executable proof, we show that affected-version ranges under-report exposure, surfacing 149 confirmed vulnerable-but-unlisted versions across 45 Python web CVEs, and that most of this exposure is invisible to replay.
  \item \textbf{An instrument for the task.} We construct \bench, a benchmark of 127 validated migration instances, each with a containerized, version-pinned environment and a vulnerability-specific oracle, admitted only after two validators confirm that the disclosed vulnerability is what triggers it.
  \item \textbf{An evaluation of automated enumeration.} We measure 18 agent configurations and a trace-guided baseline as tools for closing, or exploiting, the gap, and analyze reference-package dependence, adaptation difficulty, and the failures that remain.
\end{itemize}

\section{A Motivating Example for PoC Migration}
\label{app:motivation-example}

CVE-2024-45230~\cite{CVE-2024-45230} reports a denial-of-service vulnerability in Django's \texttt{urlize()} and \texttt{urlizetrunc()} template filters, in which a crafted input drives quadratic-time processing inside the \texttt{trim\_punctuation()} routine that strips leading and trailing punctuation from candidate URLs.
NVD lists the affected range as Django 4.2 before 4.2.16, 5.0 before 5.0.9, and 5.1 before 5.1.1.
Django 4.0 is not listed.
Running the public PoC against Django 4.0 confirms NVD's implicit claim at first glance: the crafted input is processed in linear time and the vulnerability appears not to exist.
Listing~\ref{lst:urlize-fail} shows the reference payload and its harmless outcome.
\begin{lstlisting}[caption={Reference PoC processed in linear time on Django 4.0. The trailing-semicolon branch of \texttt{trim\_punctuation()} that the payload targets does not yet exist in this version.},label={lst:urlize-fail}]
# Reference PoC, targeting Django 5.0
# Payload: "http://a&" + ";." * 100000
# Result on Django 4.0:
#   processed in 0.006s (no amplification)
#   the ';.' payload triggers no quadratic behavior.
\end{lstlisting}
The reference payload targets the semicolon-aware branch of \texttt{trim\_punctuation()}, which repeatedly re-scans the candidate string while stripping trailing \texttt{;} characters adjacent to an \texttt{\&} entity marker.
This branch, and the quadratic behavior CVE-2024-45230 fixes, was introduced in the 4.2 line; on Django 4.0 the payload falls through a simpler linear path and no amplification occurs.
Stopping here would incorrectly conclude that Django 4.0 is safe.
A deeper inspection of Django 4.0's source reveals that \texttt{trim\_punctuation()} is nonetheless quadratic, but through a different branch.
The wrapping-punctuation path, which balances leading and trailing brackets, calls \texttt{middle.count(opening)} and \texttt{middle.count(closing)} once per iteration of its trimming loop; for an input of $N$ leading brackets, the loop runs $N$ times and each iteration performs an $O(N)$ scan, yielding $O(N^2)$ total work on the same routine the patch later guards.
Migrating the PoC means abandoning the semicolon vector and constructing an input that drives the wrapping-punctuation loop instead.
Listing~\ref{lst:urlize-success} shows the migrated PoC.
\begin{lstlisting}[caption={Migrated PoC succeeds on Django 4.0. Switching from the semicolon vector to a wrapping-punctuation payload reaches the same quadratic routine through a different branch.},label={lst:urlize-success}]
# Migrated PoC, targeting Django 4.0
# Payload: "(" * 100000 + "http://a)"
# End-to-end urlize latency (same host):
#   n=10000  -> 0.09s
#   n=20000  -> 0.35s
#   n=50000  -> 2.04s
#   n=100000 -> 7.26s   (baseline: 0.004s)
# Doubling n quadruples the time: O(n^2) confirmed.
\end{lstlisting}
The migrated PoC exhibits a clean quadratic latency curve: each doubling of the input length roughly quadruples the processing time, against a $4$\,ms baseline for a benign input of the same size.
The CVE-specific oracle records success when the attack-to-baseline latency ratio on a single host exceeds a pre-registered threshold while a benign control of equal length stays below it, so the verdict depends on the amplification factor rather than on any absolute wall-clock time.
Applying the CVE-2024-45230 patch to Django 4.0's \texttt{trim\_punctuation()} collapses this curve back to linear, confirming that the migrated payload triggers the disclosed vulnerability rather than an unrelated slowdown.
Django 4.0 is therefore vulnerable to the same root cause as the NVD-listed versions, despite being outside the reported range.
We note that Django 4.0 had reached end of life before CVE-2024-45230 was disclosed, which is precisely why no advisory covers it: upstream does not assess unsupported branches, so the version is absent from every database by default rather than by verdict.
Such branches are exactly the population our study targets, which are still deployed, sharing the vulnerable code, yet invisible to the metadata every scanner consumes.

This single case exhibits the second and third challenges identified in Section~\ref{sec:intro} at once.
The reference PoC's exploitation path, which is the semicolon branch of \texttt{trim\_punctuation()}, is absent in Django 4.0, so migration cannot be a replay, and recovering the exploit requires discovering that the \emph{same} routine is reachable quadratically through a different branch.
The success oracle is an end-to-end latency ratio rather than a crash, so the crash-based validation used by prior PoC migration tools does not apply and a vulnerability-specific oracle is required.
Recovering the migration required reading Django's \texttt{urlize()} internals, identifying the alternate quadratic branch, constructing an input that drives it, and then building an oracle that distinguishes genuine amplification from hardware noise and incidental slowdowns.
Doing this at the scale of all Python web CVEs, across all prior versions, is impractical as manual work.

\section{Threat Model}
\label{sec:threat-model}

\textbf{The trusted boundary.}
Our threat model centers on the operational assumption that underpins modern vulnerability management: that a version lying outside a CVE's NVD-listed affected range is not vulnerable.
SCA scanners, automated patching pipelines, and incident-response teams all consume this boundary to decide which deployments warrant attention, so a version outside the range is effectively withdrawn from every defensive control at once.
The threat we study is the discrepancy between this recorded boundary and actual exploitability: a presumed-safe version that is, in fact, reachable by the disclosed flaw is a deployment that the entire defensive stack overlooks in unison, with no signal anywhere that contradicts the ``safe'' verdict.

\textbf{Adversary.}
We consider an adversary who already possesses a working PoC for a disclosed vulnerability on a version NVD lists as affected, and who seeks to establish that an earlier, unlisted version is exploitable by the same root cause.
The adversary can read the source of the target version, compare it against the vulnerability-fixing commit, and adapt the reference PoC through non-trivial but bounded changes such as rewriting an API entry point, tuning a payload, shifting to an alternate attack vector, or reconstructing the historical runtime environment.
The adversary does not discover new vulnerabilities; the goal is to show that a \emph{known} flaw remains reachable on a version the database treats as safe, thereby extending an n-day onto deployments outside the believed blast radius.
Each demonstrated case is a deployment that scanners pass over, patching campaigns exclude, and responders never triage.
The defender we model has strictly less information leverage than this adversary: the defender holds the same public materials but, lacking executable confirmation, is forced to trust the listed boundary. Measuring what the adversary can confirm, therefore, directly measures what the defender is blind to.

\textbf{Asset and impact.}
The assets at risk are deployed software instances running versions outside the listed affected range and therefore presumed safe.
A successful migration demonstrates that the affected-version metadata under-reports the true exposure, so the practical impact is measured directly: it is the number of unlisted-but-exploitable versions an adversary, or an automated tool acting on either side, can confirm, each of which corresponds to a class of deployments that current defenses cannot see.
Because every confirmed case is recorded as a reproducible, oracle-witnessed artifact, the impact in this paper is not asserted but counted.

\textbf{Legitimacy of a migration.}
Because the benchmark records each confirmed exposure as executable evidence, we admit a migration only when it
(i)~preserves the original vulnerability class, attack surface, and exploitation primitive;
(ii)~modifies only the PoC and its runtime environment, never the affected product's source, and in particular never reverts, monkey-patches, or wraps away the security fix;
and (iii)~is witnessed by the CVE-specific oracle and independently confirmed by two validators to fire because the disclosed vulnerability was genuinely triggered, not because of an unrelated error or a default configuration.
Outputs that satisfy the oracle by circumventing these constraints are excluded as illegitimate (Section~\ref{sec:validation}).
These criteria ensure that every counted instance reflects exposure that is genuinely present in the shipped version, which is what makes the resulting count a sound measurement of the defender's blind spot rather than an artifact of how the exploit was constructed.

\textbf{Scope.}
The in-scope ecosystems, the five-version migration window, and the explicit exclusions are stated together with the task formulation in Section~\ref{sec:task}.

\section{\bench Benchmark}
\label{sec:benchmark}

\subsection{Task Formulation}
\label{sec:task}
 
\paragraph{Setup.}
Let $c$ be a disclosed CVE affecting a Python package $p$ with release history $V(p)$. Each vulnerability database $d \in \mathcal{D} = \{\mathrm{NVD}, \mathrm{OSV}, \mathrm{GHSA}, \mathrm{Snyk}\}$ records an affected-version set $A_d(c) \subseteq V(p)$. Because different tools consume different feeds, we call a version $v$ \emph{unlisted} only when every feed excludes it, that is, $v \notin \bigcup_{d \in \mathcal{D}} A_d(c)$. This definition concerns the recorded ranges, not a guarantee of safety. Let $v_{\min}(c)$ be the earliest version any feed lists as affected. Following the pilot study in Section~\ref{sec:pilot}, the \emph{candidate versions} of $c$ are the five most recent releases strictly preceding $v_{\min}(c)$, each of which is unlisted by construction.
 
\paragraph{Task.}
Given a reference PoC $P$ validated to trigger $c$ on a listed version $v_{\mathrm{ref}}$, and a candidate version $v$, cross-version PoC migration asks for a modified PoC $P'$ that triggers the same vulnerability on $v$. A benchmark instance is a tuple $(c, v, E, P, O)$, where $E$ is a container pinned to $v$, $P$ is the reference PoC in the standardized specification of Appendix~\ref{app:reference-details}, and $O$ is a deterministic oracle declared for $c$. A solver additionally receives the CVE advisory, the vulnerability-fixing commit, and the source tree of $v$, and may execute commands, inspect logs, and edit PoC files inside $E$.
 
\paragraph{Success criterion.}
An output $P'$ is accepted only after passing two gates. The first is the oracle. Executing $P'$ against $E$ must satisfy $O$, which is declared per CVE because web vulnerabilities have no universal success signal (Section~\ref{sec:validation}). A SQL injection oracle checks the response body for a pre-planted canary row, whereas a ReDoS oracle compares attack latency against a benign control input of equal length. The second gate is causal validity. Two validators must independently agree that $O$ fired because $c$ was genuinely triggered on $v$, rather than for an incidental reason such as an unrelated error or a default configuration.
 
\paragraph{Constraints on a valid migration.}
A migration counts only if it (i) preserves the vulnerability class, attack surface, and exploitation primitive of $c$, so substituting a weaker effect such as denial of service for remote code execution is rejected; (ii) modifies the PoC and its runtime environment but never the source of the affected package, and in particular never reverts, monkey-patches, or wraps away the security fix; and (iii) satisfies both gates above. Together these ensure that a resolved instance reflects exposure present in the shipped version rather than an artifact of how the exploit was constructed. Outputs that satisfy $O$ by circumventing them are recorded as illegitimate (Section~\ref{sec:rq3}).
 
\paragraph{Interpretation.}
A resolved instance provides executable evidence that an unlisted version is affected by the disclosed vulnerability. It does not establish how many deployments run that version or how individual security tools classify it.

\paragraph{Scope.}
We include disclosed CVEs in the Python web ecosystem (Django, Flask, FastAPI, Web2py, and applications built on them).
They must have a public patch commit and sufficient context to reconstruct a PoC, together with migration to the five candidate versions defined above. 
We treat the following as out of scope: previously unknown (0-day) vulnerabilities, over-reporting (versions incorrectly listed as affected), ecosystems beyond Python web, and exposure on versions outside the five-version window.
Our counts are therefore a lower bound on under-reporting within that window rather than a prevalence estimate over full release histories.

\subsection{Comparison with Existing Benchmarks}
\label{app:comparison}

Table~\ref{tab:bench-compare} positions \bench against representative cybersecurity and coding benchmarks for AI agents and against prior cross-version exploit-migration work. Unlike these benchmarks, \bench starts from a validated reference PoC and requires adapting it to an earlier, unlisted version: success is witnessed by a vulnerability-specific oracle rather than a flag, crash, or repository test, and every resolved instance is executable evidence of an under-reported vulnerable version. Section~\ref{sec:related-work} discusses these benchmarks in more detail.

\begin{table}[!htbp]

\centering
  \renewcommand{\arraystretch}{1.0}
  \captionsetup{font=small,skip=4pt}
\footnotesize
\caption{Representative benchmarks compared with \bench, grouped by task family. Counts are as reported by each primary source; ``--'' means not reported.}
\label{tab:bench-compare}
\setlength{\tabcolsep}{3pt}
\begin{tabular}{lllrr}
\toprule
\textbf{Benchmark} & \textbf{Domain} & \textbf{Success oracle} & \textbf{\# Inst.} & \textbf{\# Proj.} \\
\midrule
\multicolumn{5}{l}{\emph{Cybersecurity benchmarks for AI agents}} \\
Cybench \cite{zhang2025cybench}               & CTF              & Flag string          & 40      & --  \\
CyberGym \cite{wang2026cybergym}              & C/C++ (OSS-Fuzz) & Crash, dual-exec     & 1{,}507 & 188 \\
CVE-Bench \cite{zhu2025cvebench}              & Web apps         & 8 std.\ attack goals & 40      & 26  \\
\midrule
\multicolumn{5}{l}{\emph{Coding benchmarks for AI agents}} \\
SWE-bench \cite{jimenez2024swe}          & Python           & Repo tests (F$\to$P) & 2{,}294 & 12  \\
SWT-bench \cite{muendler2024swtbench}         & Python           & Repo tests (F$\to$P) & 433     & --  \\
GitChameleon 2.0 \cite{misra2026gitchameleon} & Python libs      & Unit tests           & 328     & 26  \\
\midrule
\multicolumn{5}{l}{\emph{Cross-version exploit migration}} \\
VulScope \cite{vulscope}                      & C/C++ binaries   & Crash + triage       & 470\textsuperscript{$\ddagger$} & 6 \\
Diffploit \cite{chen2025diffploit}            & Java libs        & Executable repro.    & 689     & 79 \\

\midrule
\textbf{\bench (ours)} & \textbf{Python web} & \textbf{Per-CVE oracle} & \textbf{127}\textsuperscript{$\dagger$} & \textbf{17} \\
\bottomrule
\end{tabular}
\\[2pt]
{\footnotesize \textsuperscript{$\dagger$}127 migration instances across 45 CVEs, plus 22 prior versions on which the reference PoC works unchanged.\quad
\textsuperscript{$\ddagger$}470 target CVE-version pairs derived from 30 reference CVEs.}

\end{table}

\subsection{Pilot Study}
\label{sec:pilot}

The pilot calibrates the number of prior versions to examine and the time budget for each manual migration. Exhaustive evaluation is costly because packages can have hundreds of releases, and additional effort does not guarantee a successful migration.

\textit{Process.} We randomly sampled 10 CVEs across the in-scope frameworks, balancing CWE categories, with at most two CVEs from any single CWE. 
For each pilot CVE, we attempted to migrate the PoC to every available prior version on PyPI, yielding an average of 14 versions per CVE and 136 attempts in total.

For each attempt, we recorded the wall-clock time and assigned one of four outcomes.
(1) \textit{Successfully migrated} marks attempts in which an adapted PoC triggers the vulnerability on the prior version. 
(2) \textit{No adaptation} marks attempts in which the unmodified reference PoC already triggers it. 
(3) \textit{Migratable failure} marks attempts in which the vulnerable code path appears reachable but neither author could produce a working exploit within the attempt budget. 
(4) \textit{Not affected} requires two conditions together. The first is static: a manual inspection of the prior version's source against the patch commit indicates that the vulnerable code path is absent or structurally distinct from the disclosed version. The second is dynamic: both authors fail to trigger the vulnerability within an extended three-hour budget. 
We require both criteria because each guards against a different misclassification. Source inspection and execution provide complementary evidence, although neither establishes the absence of every possible exploitation path.

\begin{figure}[t]
  \centering
  \begin{subfigure}[b]{0.48\textwidth}
    \includegraphics[width=\linewidth]{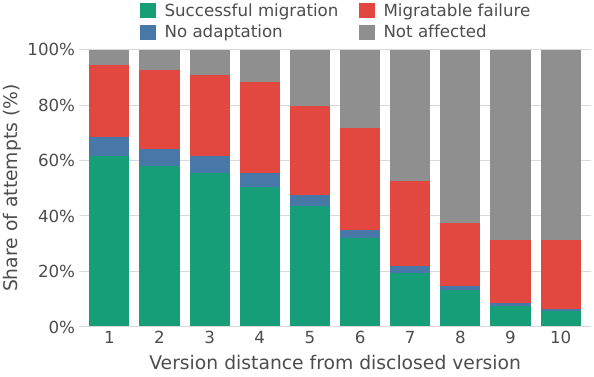}
    \caption{Outcome distribution by version distance.}
    \label{fig:pilot_distance}
  \end{subfigure}
  \hfill
  \begin{subfigure}[b]{0.48\textwidth}
    \includegraphics[width=\linewidth]{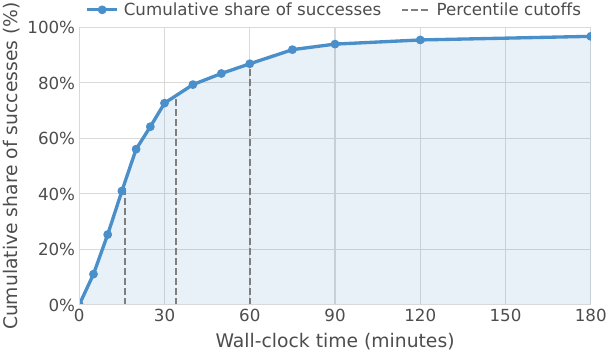}
    \caption{Cumulative time-to-success.}
    \label{fig:pilot_time}
  \end{subfigure}
  \caption{Pilot study results across 136 migration attempts on 10 CVEs.}
  \Description{Two bar charts summarising pilot migration success by version distance and time-to-success.}
  \label{fig:pilot}
\end{figure}

Figure~\ref{fig:pilot_distance} shows that successful migrations become less frequent with version distance, from roughly 62\% at distance 1 to 6\% at distance 10. Beyond distance 5, most examined versions are classified as \textit{not affected}. We therefore examine the five most recent prior versions per CVE.

Among successful pilot migrations, the median, 75th percentile, and 90th percentile of time-to-success are approximately 17, 35, and 60 minutes (Figure~\ref{fig:pilot_time}). We set the per-attempt budget to 60 minutes, which covers approximately 90\% of the observed pilot successes. This choice bounds annotation effort but can exclude migrations that require more time.

\subsection{Benchmark Construction}

\begin{figure*}[t]
\centering
\includegraphics[width=\textwidth]{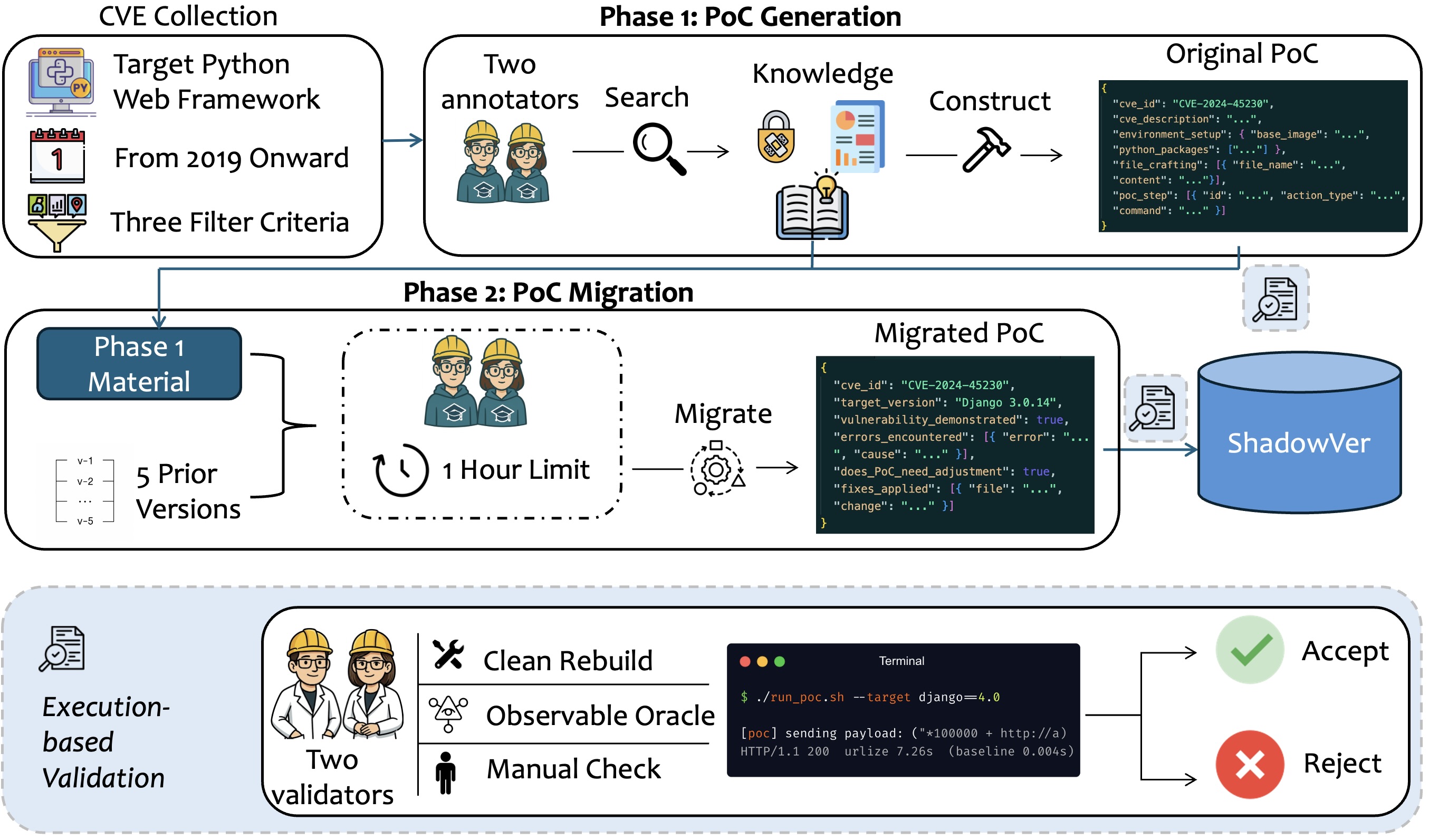}
\caption{Overview of \bench construction pipeline}
\Description{Pipeline diagram of the benchmark construction: CVE collection, reference PoC generation, cross-version PoC migration, and validation.}
\label{fig:pipeline}
\end{figure*}


\subsubsection{CVE Collection and Scope}
\label{sec:cve-collection}
We study Python web frameworks (Django, Flask, FastAPI, and Web2py), Python applications, and related packages such as urllib3, Plone, and aiohttp.

We systematically collect CVEs published from January 2019 onward from the NVD.
We include a CVE if it satisfies three criteria: 
(i)~the vulnerability affects a Python web framework or application within our scope; 
(ii)~a publicly accessible patch commit is available, enabling root-cause analysis; 
and (iii)~sufficient contextual evidence, such as advisory descriptions, issue-tracker discussions, and prior writeups, exists to support PoC development.
Applying these criteria yields an initial set of 217 CVEs.
We also query the four feeds, NVD, OSV, GHSA, and Snyk, to check whether any version we later confirm vulnerable has been recorded as affected.
A version counts as unlisted only when all four exclude it, under our cross-database definition.

\subsubsection{Execution-Based Validation}
\label{sec:validation}
Each candidate from either construction phase is checked by two validators. They verify that the environment can be rebuilt, the declared oracle passes, and the observed effect is caused by the disclosed vulnerability. All reported accepted-instance counts are post-validation counts.

Web vulnerabilities require different execution checks. Unlike crash-based checks used for many memory-corruption tasks~\cite{vulscope}, SQL injection, XSS, path traversal, and resource exhaustion have different observable effects. We therefore specify a programmatically checkable oracle for each PoC. Oracle patterns are given in Appendix~\ref{app:oracle-patterns}.

Instance-specific execution checks are also used in SecBench.js~\cite{bhuiyan2023secbenchjs} and SWE-bench~\cite{jimenez2024swe}. Each validator independently rebuilds the environment on a separate machine, executes the recorded steps, and evaluates the oracle. Inconsistent outcomes prompt revisions to the environment or specification and another validation pass.

The validators also inspect the relevant requests, responses, logs, program state, and target source to determine whether the disclosed vulnerability caused the observed effect. Initial assessments are recorded independently; disagreements are then resolved through discussion. Cohen's kappa is 0.78 for Phase~1 and 0.81 for Phase~2. Each accepted instance includes a version-pinned Docker image, a JSON PoC specification with its oracle, and a script that provisions the environment, executes the PoC, and reports the verdict.

\subsubsection{Phase 1: PoC Generation}
\label{sec:phase1}

Our benchmark requires a complete, runnable PoC for every CVE, but public exploits are scarce or fragmented for most of them and rarely execute as published~\cite{householder2020historical, mu2018understanding, lee2026secbench}.
We therefore construct each PoC from scratch, with two annotators developing and cross-reviewing each one.
Construction details and the standardized PoC specification are provided in Appendix~\ref{app:reference-details}.

Across all candidates, the two annotators invested approximately 242 person-hours in PoC construction. 
A candidate was carried forward only if the annotators could produce a PoC that subsequently passed the validation protocol of Section~\ref{sec:validation}.
This construction process began with 217 candidate CVEs and, after validation, yielded 97 accepted PoCs, as 120 candidates were filtered out.
Specifically, 47 candidates provided insufficient data for reconstruction, 33 involved deprecated dependencies, and 11 required unobtainable proprietary software.
An additional 29 cases failed during build or runtime.

\subsubsection{Phase 2: Cross-version PoC Migration}
\label{sec:phase2}

For each of the 97 validated PoCs from Phase~1, we attempt to migrate the exploit to prior versions that the corresponding CVE does not list as affected.
 
Guided by the pilot study (Section~\ref{sec:pilot}), we target the five most recent versions preceding the earliest affected version across four feeds, which we call the ``candidate versions'' for the CVE.
For each candidate version, both authors independently attempt to adapt the reference PoC so that it triggers the same vulnerability on the target version. The two work on separate machines with no shared notes during the attempt, each under the 60-minute budget calibrated in the pilot, and starting from the same materials: the CVE advisory, the patch commit, the reference PoC, and the public sources used in Phase~1. Each annotator invested approximately 291 hours across the candidate-version attempts.
Canonical PoC selection and the six screening outcomes are detailed in Section~\ref{app:migration-details}.

Across the 97 Phase~1 CVEs, we examined 297 candidate versions (Table~\ref{tab:ledger}, Section~\ref{app:migration-details}). We confirmed 149 vulnerable versions: 127 require migration and form the benchmark, while 22 work with the unchanged reference PoC. The remaining 148 comprise 61 classified as \textit{not affected} and 87 inconclusive outcomes. Inconclusive outcomes do not establish the absence of the vulnerability.

A CVE enters the benchmark if at least one candidate version yields a validated migration. This gives 127 instances across 45 CVEs. Among the other 52 CVEs, 23 have all examined candidates classified as \textit{not affected}, while 29 have at least one inconclusive candidate and no confirmed vulnerable candidate. Their outcomes remain in the screening ledger.

\subsubsection{Migration Categories}
\label{sec:migration_categories}

To characterize how the migrated PoCs differ from their reference PoCs, we define five categories of adaptation, assigned during Phase~2; each migrated PoC carries one or more labels.
\textbf{API Import Change (58):} the migrated PoC imports a different module, symbol, or entry point than the reference, because the original was renamed, moved, or modified between versions.
\textbf{Payload Tuning (41):} the payload differs in size, encoding, character set, or repetition count, to bypass a version-specific filter or parser or to match a version-specific input-handling quirk.
\textbf{Attack Vector Shift (29):} the PoC reaches the same flaw through a different exploitation path, e.g., a different sink leading to the vulnerable code.
\textbf{Environment Dependency Setup (22):} the runtime environment differs in Docker base image, Python interpreter version, database backend, or other transitive dependencies.
\textbf{Config Change (17):} configuration files, templates, or deployment settings differ, as needed to instantiate the vulnerable code path on the target version.
A migrated PoC may carry more than one label, so these counts sum to more than the 127 instances. The most common adaptations involve the API surface and the exploit payload, while attack-vector shifts, environment repairs, and configuration changes are less frequent. As later sections show, frequency does not track difficulty: attack-vector shifts appear in fewer than a quarter of instances yet are the hardest category overall.

\subsubsection{Migration Adjudication and Outcome Definitions}
\label{app:migration-details}

If both independent attempts succeed, the two authors reconcile their results to synthesize a single canonical migrated PoC. This version is either selected from the individual attempts or derived by combining their most robust elements. When the attempts target distinct code paths, priority is given to the version that demonstrates higher fidelity to the root cause detailed in the CVE advisory. If only one attempt succeeds, it is automatically designated as canonical. All canonical migrated PoCs must subsequently pass the validation protocol described in Section~\ref{sec:validation}. Instances that fail to yield a reproducible migrated PoC within the allocated budget are recorded as failures and excluded from the dataset.

During manual migration, we recorded one of six outcomes for each candidate version. Four reuse the pilot definitions (Section~\ref{sec:pilot}): \textit{successful migration}, \textit{no adaptation}, \textit{migratable failure}, and \textit{not affected}. 
  
The remaining two refine outcomes the pilot recorded more coarsely. (5) \textit{Timeout} marks attempts in which the vulnerable code path appears reachable but neither author reached a working exploit within the 60-minute budget; the pilot folded these into \textit{migratable failure}. (6) \textit{Unbuildable} marks candidate versions whose historical environment could not be built, so the PoC was never executed against them; the pilot folded these into \textit{not affected}. We separate \textit{unbuildable} from \textit{not affected} because the two rest on different observations: a \textit{not affected} verdict follows from inspecting the target source and confirming the vulnerable code is absent, whereas an \textit{unbuildable} version was never tested at all.

\begin{table}[!htbp]
  \centering
  \setlength{\tabcolsep}{4pt}
  \renewcommand{\arraystretch}{1.0}
  \captionsetup{font=small,skip=4pt}
  \caption{Phase 2 version-level screening ledger: outcomes for all 297 candidate versions examined across the 97 validated CVEs.}
  \label{tab:ledger}
  \footnotesize
  \begin{tabular}{llrr}
    \toprule
    Evidence & Outcome & Versions & Share \\
    \midrule
    \multirow{2}{*}{Confirmed vulnerable}
      & Successful migration & 127 & 42.8\% \\
      & No adaptation        &  22 &  7.4\% \\
    \cmidrule(lr){2-4}
      & \textit{Subtotal}    & \textit{149} & \textit{50.2\%} \\
    \midrule
    Not affected by the CVE
      & Not affected         &  61 & 20.5\% \\
    \midrule
    \multirow{3}{*}{Inconclusive}
      & Unbuildable          &  34 & 11.4\% \\
      & Migratable failure   &  27 &  9.1\% \\
      & Timeout              &  26 &  8.8\% \\
    \cmidrule(lr){2-4}
      & \textit{Subtotal}    &  \textit{87} & \textit{29.3\%} \\
    \midrule
    & \textbf{Total}         & \textbf{297} & \textbf{100.0\%} \\
    \bottomrule
  \end{tabular}
\end{table}

\section{Study Design}
Our evaluation is organized around the following three research questions (RQs):
\begin{itemize}
    \item \textbf{RQ1:} What are the characteristics of PoC migration instances in the Python web ecosystem?
    \item \textbf{RQ2:} How do existing and LLM-based migration techniques perform on \bench?
    \item \textbf{RQ3:} Why do current methods fail, and which properties predict failure?
\end{itemize}

\subsection{Methodology for RQ1}
\label{sec:method-rq1}
RQ1 characterizes the benchmark along three aspects: the prevalence of under-reported vulnerable versions, the severity and type of the vulnerabilities, and the types of adaptation each migration requires.  

We measure under-reporting prevalence at the version level rather than the CVE level. For each CVE, we count the prior versions within the five-version window calibrated in the pilot (Section~\ref{sec:pilot}) that were confirmed vulnerable through executable PoCs. We distinguish two cases: versions exploitable by the unmodified reference PoC (Phase~2 outcome 1), and versions exploitable only after migration (Phase~2 outcome 2). The sum is our count of confirmed vulnerable-but-unlisted versions. 
This is a count over the prior versions we examined and validated, not a base rate over all prior versions, which would require auditing every version of every package in the ecosystem and is outside the scope of this work.

We characterize severity and vulnerability type using values published in the corresponding NVD entry for each of the 45 CVEs. 
CVSS scores are taken as the v3 base score and bucketed into the standard severity bands. CWE assignments are also taken from NVD as published. Two CVEs are listed by NVD as NVD-CWE-noinfo, meaning no CWE was assigned by the cataloging authority. We report these in their own bucket rather than reassign them.

For adaptation effort, we report the distribution at the level of migration instances. For each label defined in Section~\ref{sec:migration_categories}, we count how many of the 127 instances carry it. Because a single instance can carry more than one label, these counts sum to more than 127.

\subsection{Methodology for RQ2}
\label{sec:method-rq2}
We evaluate three families of approaches. 
The first is an adapted version of VulScope~\cite{vulscope}, the state-of-the-art trace-guided migration tool originally designed for C/C++ userspace binaries. We describe the adaptation below and refer to the result as \vsw. 
The second and third are two widely used agent harnesses, OpenHands~\cite{openhands} and SWE-agent~\cite{sweagent}. 
We instantiate each harness with nine backbone models spanning a range of capabilities: GPT-OSS-120b~\cite{gpt_oss}, Qwen-3.5~\cite{qwen2025qwen3}, Kimi-2.5~\cite{kimi2026k25}, GLM-5.2~\cite{zhipu2026glm52}, GPT-5.3~\cite{openai2026gpt53}, Grok-4.6~\cite{xai2026grok46}, Gemini-3.1~\cite{googledeepmind2026gemini31}, Sonnet-4.6~\cite{anthropic2026sonnet46}, and Opus-4.6~\cite{anthropic2026opus46}. 
This gives 18 agent configurations plus the \vsw baseline, 19 tool configurations in total.
The baseline adaptation is described in Appendix~\ref{app:baseline-adaptation}, with implementation details in Appendix~\ref{app:vulscopeweb}.

Crucially, \vsw preserves VulScope's input-only action space: it mutates only the exploit input exposed by the reference PoC (HTTP method, path, query, headers, cookies, form and JSON fields, uploaded content, and declared payload parameters) and never edits the harness code, imports, Dockerfile, dependencies, configuration, or product source. 
The LLM agents we evaluate have a strictly broader action space, since they may also edit the PoC and its environment. 
We therefore use \vsw to test whether VulScope's central assumption still holds for Python web migration, namely, whether a target vulnerability can be reached by adjusting inputs while keeping the execution path similar to the reference trace. 
When a migration instead requires changing APIs, switching attack vectors, repairing a historical environment, or modifying configuration, \vsw is structurally unable to perform it, and these failures identify exactly where the input-adjustment premise behind trace-guided migration breaks down.

\textit{Run protocol.} Each tool receives the reference PoC produced in Phase 1, the source tree of the target version, and a containerized environment matching the target version. The agent configurations have access to a terminal and a file editor inside this environment, with which they can execute commands, inspect logs, edit the PoC files, and observe outcomes. 
Each agent run is bounded by a budget of 10M tokens, counted inclusively over input, output, and reasoning tokens for the target model and excluding oracle tokens (see Appendix~\ref{app:budget} for the budget rationale). 
Each tool was run once on each of the 127 migration instances.

Each configuration is evaluated once per instance, so run-to-run variance is unmeasured (Section~\ref{sec:threats}). Small differences should be interpreted cautiously. Across the 19 tool configurations, the total LLM API cost of the experiment was approximately \$926.
 
\textit{Agent task specification.} Agents are instructed to preserve the vulnerability class, attack surface, and exploitation primitive, and to modify only the PoC and its environment. They may not edit the affected product's source or circumvent its security fix. If a valid migration cannot be found, they must report failure. These are the task constraints in Section~\ref{sec:task}.

\textit{Acceptance criterion.} Agent outputs undergo the oracle and causal-validity checks in Section~\ref{sec:validation}. The validators who constructed the reference instances review these outputs while blinded to the producing model and harness, then reconcile disagreements. Outputs that rely on modifying the affected product are rejected. For \vsw, the evaluation uses the oracle alone because the baseline changes only inputs against a fixed entry point. This difference in validation is a limitation of the comparison.

\textit{Validator effort and agreement.} Reviewing each (instance, tool) pair took 14.2 minutes on average. Inter-validator agreement on the binary accept-or-reject decision was Cohen's kappa = 0.83, indicating almost perfect agreement~\cite{landis1977measurement}. 
The pairs that both validators rejected form the failure pool analyzed in Section~\ref{sec:rq3}, following the protocol in Section~\ref{sec:method-rq3}.
 
\textit{Metrics.} We report two views of the resolve rate. The aggregate view is the share of the 127 instances each tool resolves. The per-category view, for each of the five migration categories defined in Section~\ref{sec:migration_categories}, is the share of instances carrying that label that the tool resolves. Since a single instance may carry more than one label, the per-category sample sizes do not partition the benchmark. Each cell is computed over all instances bearing the column's label, regardless of co-occurring labels.

\subsection{Methodology for RQ3}
\label{sec:method-rq3}
 
We analyze failed runs to distinguish problems in selecting an adaptation from problems in completing and verifying it.
We combine a qualitative categorization of failure root causes with a correlation analysis of category shares against benchmark resolve rate. 
The detailed coding and correlation protocol is provided in Appendix~\ref{app:failure-protocol}.

\section{Evaluation Results}
\label{sec:evaluation}

\subsection{RQ1: What are the characteristics of PoC migration instances in the Python web ecosystem?}
\label{sec:rq1}

Before evaluating automated tools, we characterize the migration instances in the benchmark and what they reveal about under-reported vulnerability exposure.
We describe the instances along two axes here, the relationship between versions that require migration and those that do not, and the severity and type of the vulnerabilities involved; the adaptation each migration requires is characterized with the benchmark construction (Section~\ref{sec:migration_categories}).

Of the 149 confirmed vulnerable prior versions, 22 (14.8\%) work with the unchanged reference PoC and 127 (85.2\%) require adaptation. The former are excluded from migration evaluation. These counts concern the examined versions and exclude inconclusive outcomes; they do not estimate ecosystem-wide prevalence.
The vulnerability profile is provided in Section~\ref{app:profile}.

\subsubsection{Vulnerability Profile}
\label{app:profile}

We characterize the benchmark by the need for adaptation, vulnerability severity, and CWE category.

\begin{figure}[!t]
  \centering
  \begin{subfigure}[b]{0.48\columnwidth}
    \centering
    \includegraphics[width=\linewidth]{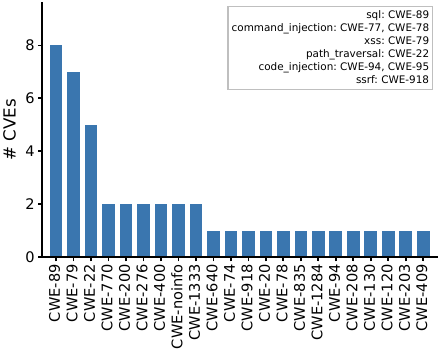}
    \caption{CWE distribution}
    \label{fig:cwe_dist}
  \end{subfigure}
  \hfill
  \begin{subfigure}[b]{0.48\columnwidth}
    \centering
    \includegraphics[width=\linewidth]{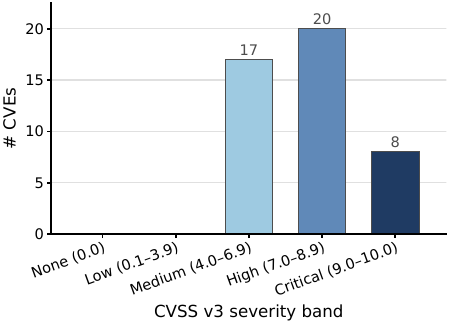}
    \caption{CVSS v3 severity distribution}
    \label{fig:severity_dist}
  \end{subfigure}
  \caption{Vulnerability profile of the 45 benchmark CVEs.}
  \Description{Two charts showing the CWE distribution and CVSS severity distribution of the 45 benchmark CVEs.}
  \label{fig:profile}
  \vspace{-1em}
\end{figure}

The benchmark covers a range of typical web-application vulnerabilities, with most CVEs carrying substantial security impact.
Of the 45 CVEs, 28 (62\%) are rated High or Critical under CVSS v3, and none fall below Medium (Figure~\ref{fig:severity_dist}). 
Figure~\ref{fig:cwe_dist} shows the CWE distribution: CWE-89 (SQL injection, 8 CVEs), CWE-79 (cross-site scripting, 7), and CWE-22 (path traversal, 5), which together account for 44\% of the benchmark, followed by a long tail that includes resource exhaustion (CWE-400, CWE-770), server-side request forgery (CWE-918), and code injection (CWE-94).

\find{\textbf{Answer to RQ1:} Across the 45 CVEs, 149 prior versions are exploitable despite being absent from the listed affected range. Only 22 of them fall to the unmodified reference PoC; the remaining 127 require non-trivial migration, which is exposure that replay alone cannot reveal.}

\subsection{RQ2: How do existing and LLM-based migration techniques perform on \bench?}
\label{sec:rq2}

\begin{table*}[!htbp]
  \centering
  \setlength{\tabcolsep}{0pt}
  \renewcommand{\arraystretch}{1.1}
  \captionsetup{font=small,skip=4pt}
  \caption{Resolve rates on the 127 migration instances. Each cell reports resolved
           instances out of 127, with the resolve rate in parentheses.
           \emph{\vsw} resolves 11/127 (8.7\%).}
  \label{tab:rq2-aggregate}
  \scriptsize
  \begin{tabular*}{\textwidth}{@{\extracolsep{\fill}}l *{9}{c}@{}}
    \toprule
    Agent & GPT-OSS & Qwen-3.5 & Kimi-2.5 & GLM-5.2 & GPT-5.3
          & Grok-4.6 & Gemini-3.1 & Sonnet-4.6 & Opus-4.6 \\
    \midrule
    OpenHands & 17 (13.4\%) & 31 (24.4\%) & 33 (26.0\%) & 56 (44.1\%) & 57 (44.9\%)
              & 60 (47.2\%) & 61 (48.0\%) & 64 (50.4\%) & \textbf{74 (58.3\%)} \\
    SWE-agent & 13 (10.2\%) & 30 (23.6\%) & 33 (26.0\%) & 54 (42.5\%) & 55 (43.3\%)
              & 59 (46.5\%) & 61 (48.0\%) & 63 (49.6\%) & \textbf{71 (55.9\%)} \\
    \bottomrule
  \end{tabular*}
\end{table*}

Table~\ref{tab:rq2-aggregate} reports aggregate resolve rates. \vsw resolves 11 of 127 instances (8.7\%), while agent configurations range from 10.2\% to 58.3\%. The best result is 74 instances with Opus-4.6 in OpenHands. These results leave 53 instances unresolved by that configuration; they do not establish a performance ceiling.

Differences across models are larger than differences across the two harnesses in this evaluation. For a fixed model, the harness gap is at most 3.2 percentage points. Within OpenHands, the difference between GPT-OSS-120b and Opus-4.6 is 44.9 points. Since each configuration was run once, we do not infer a definitive harness ranking or the effect of further harness tuning.

The baseline comparison also reflects different action spaces. \vsw can change inputs but cannot repair imports, environments, or configuration. Its aggregate score therefore measures this particular adaptation of trace-guided migration, rather than isolating the effect of trace guidance or oracle design.

\begin{table}[!htbp]
  \centering
  \captionsetup{font=small,skip=4pt}
  \caption{Resolve rate (\%) by migration category, per agent harness
           (OH = OpenHands, SWE = SWE-agent). Best result per column in \textbf{bold}.}
  \label{tab:rq2-bycategory}
  \scriptsize
  \setlength{\tabcolsep}{0pt}
  \renewcommand{\arraystretch}{1.05}
  \begin{tabular*}{\columnwidth}{@{\extracolsep{\fill}}l *{10}{c}@{}}
    \toprule
    & \multicolumn{2}{c}{AI (58)} & \multicolumn{2}{c}{PT (41)}
    & \multicolumn{2}{c}{AV (29)} & \multicolumn{2}{c}{ED (22)}
    & \multicolumn{2}{c}{CC (17)} \\
    \cmidrule(lr){2-3}\cmidrule(lr){4-5}\cmidrule(lr){6-7}\cmidrule(lr){8-9}\cmidrule(lr){10-11}
    Model & OH & SWE & OH & SWE & OH & SWE & OH & SWE & OH & SWE \\
    \midrule
    GPT-OSS-120b & 13.8 & 17.2 & 14.6 &  2.4 & 13.8 &  6.9 &  9.1 &  9.1 & 11.8 & 11.8 \\
    Qwen-3.5     & 36.2 & 27.6 & 31.7 & 12.2 &  6.9 & 10.3 &  4.5 & 27.3 & 23.5 & 52.9 \\
    Kimi-2.5     & 37.9 & 27.6 & 14.6 & 19.5 &  3.4 & 20.7 & 31.8 & 36.4 & 41.2 & 29.4 \\
    GLM-5.2      & 60.3 & 67.2 & 39.0 & 24.4 & 20.7 & 20.7 & 40.9 & 40.9 & 41.2 & 47.1 \\
    GPT-5.3      & 56.9 & 70.7 & 43.9 & 22.0 & 20.7 & 13.8 & \textbf{54.5} & 36.4 & 17.6 & 47.1 \\
    Grok-4.6     & 60.3 & 70.7 & 43.9 & 26.8 & 24.1 & 24.1 & 45.5 & 40.9 & 52.9 & 52.9 \\
    Gemini-3.1   & 58.6 & 58.6 & \textbf{46.3} & \textbf{31.7} & 27.6 & 34.5 & 40.9 & 45.5 & 47.1 & 70.6 \\
    Sonnet-4.6   & 72.4 & \textbf{84.5} & 29.3 & 14.6 & 17.2 & \textbf{37.9} & 36.4 & 36.4 & \textbf{88.2} & 58.8 \\
    Opus-4.6     & \textbf{79.3} & 82.8 & 43.9 & 24.4 & \textbf{31.0} & 34.5 & \textbf{54.5} & \textbf{50.0} & 70.6 & \textbf{82.4} \\
    \midrule
    \emph{\vsw}  & \multicolumn{2}{c}{0.0} & \multicolumn{2}{c}{26.8}
                 & \multicolumn{2}{c}{0.0} & \multicolumn{2}{c}{0.0}
                 & \multicolumn{2}{c}{0.0} \\
    \bottomrule
  \end{tabular*}
  \par\vspace{2pt}
  {\scriptsize\raggedright Each cell is the percentage of instances carrying the
  column's label that the row's tool resolves; labels are not mutually exclusive,
  so per-label sizes (in parentheses) sum above 127. AI~=~API Import,
  PT~=~Payload Tuning, AV~=~Attack Vector Shift, ED~=~Env Dep.\ Setup,
  CC~=~Config Change.\par}
\end{table}

Table~\ref{tab:rq2-bycategory} breaks the resolve rate down by migration category. 
Because a single instance may carry multiple category labels, each cell reports the resolve rate over all instances carrying the given label, regardless of co-occurring labels. 
The header for each column reports the number of labeled instances.

Attack Vector Shift is the hardest category overall, and the only one on which no tool configuration exceeds 40\%.
Its highest cell is 37.9\%, while every other category reaches at least 46\% for some configuration and two of them pass 70\%.
The two strongest configurations show the size of the gap: Opus-4.6 resolves 79.3\% and 82.8\% of API Import Change instances on OpenHands and SWE-agent respectively, but only 31.0\% and 34.5\% of Attack Vector Shift instances.
Recall that Attack Vector Shift instances are those in which the reference PoC's exploitation path no longer reaches the vulnerable code in the target version, requiring the migration to discover an alternate code path triggering the same root cause.
These instances require the agent to identify an alternative route to the same vulnerable behavior rather than only update local API calls or payloads. The lower resolve rates identify this form of adaptation as a challenge for the evaluated agents, although the overlapping category labels do not isolate its difficulty from other required changes.

Per-category rankings differ from the aggregate ordering. For example, Sonnet-4.6 reaches 84.5\% on API Import Change with SWE-agent but only 14.6\% on Payload Tuning. With 41 payload-tuning instances and one run per configuration, we cannot determine whether this reflects a stable model characteristic or run-to-run variation.

\vsw resolves 26.8\% of Payload Tuning instances and none in the other categories, consistent with its input-only action space. Category comparisons have small denominators: Config Change contains 17 instances and Environment Dependency Setup contains 22. Repeated runs are needed to assess the stability of these rates.

\noindent\textbf{Contribution of the reference package.}
To examine how much the supplied reference supports adaptation, we withhold the reference PoC and its environment specification from OpenHands with Sonnet-4.6. Resolve rate falls from 50.4\% (64/127) to 15.0\% (19/127), a 35.4-percentage-point decrease under the same token budget. The agent still receives the advisory, patch commit, target source, target-version container, and oracle. This result shows the value of the reference package for this configuration; it does not isolate exploit reasoning from environment reconstruction. Section~\ref{app:no-ref-poc} gives the protocol and limitations.

\subsubsection{Ablation: Migration Without a Reference PoC}
\label{app:no-ref-poc}
Each \bench instance supplies a validated reference PoC, including its environment specification. We examine how much this package supports an agent compared with the remaining task materials alone.

We re-run OpenHands with Sonnet-4.6, a high-performing configuration in Table~\ref{tab:rq2-aggregate}, on all 127 instances with the Phase~1 reference PoC and its environment specification withheld. The agent still receives the CVE advisory, patch commit, target source tree, a container matching the target version, and the instance's success oracle. The token budget, run count, and two-gate acceptance criterion of Section~\ref{sec:method-rq2} are unchanged.

Resolve rate falls from 50.4\% (64/127) to 15.0\% (19/127), a decrease of 35.4 percentage points (Table~\ref{tab:no-ref-poc}). The supplied reference package therefore substantially supports successful adaptation in this configuration. The 19 successes without it show that some instances can also be solved from the remaining materials within the same budget.

\begin{table}[h]
\centering
\caption{Effect of withholding the reference PoC. OpenHands with Sonnet-4.6 on
all 127 migration instances.}
\label{tab:no-ref-poc}
\begin{tabular}{lcc}
\toprule
Configuration & Resolved & Resolve rate \\
\midrule
Reference package provided & 64/127 & 50.4\% \\
Reference package withheld         & 19/127 & 15.0\% \\
\bottomrule
\end{tabular}
\end{table}

This comparison removes both the reference exploit and its environment specification, so it does not isolate the contribution of exploit-level information from support for environment reconstruction. It also uses one run of one agent--model configuration and does not establish the same effect for other configurations or budgets. Isolating these contributions would require retaining the environment specification while withholding only the exploit.

\find{\textbf{Answer to RQ2:} The best agent configuration resolves 58.3\% of the 127 instances, against 8.7\% for the adapted trace-guided baseline. Attack-vector shifts are the hardest category, with no configuration above 37.9\%, and in one configuration withholding the reference PoC and its environment specification lowers the resolve rate from 50.4\% to 15.0\%.}

\subsection{RQ3: Why do current methods fail, and which properties predict failure?}
\label{sec:rq3}
 
To understand why current tools leave more than 40\% of the benchmark unresolved, we analyzed the failures produced during tool evaluation. The set of failed pairs is exactly the output of the validation procedure in Section~\ref{sec:method-rq2}: any pair that the two validators marked as not resolved.

The five root-cause categories defined using the protocol in Appendix~\ref{app:failure-protocol} are:
\emph{Missed API Migration}: the agent attacked the wrong entry point in the target version, either by replaying the reference PoC's entry point or by selecting an alternative that does not reach the vulnerable sink.
\emph{Correct Strategy but Incomplete}: the agent found the correct API and strategy, but the migrated PoC never triggered the vulnerability.
\emph{Oracle Misinterpretation}: the agent declared success on output that does not exploit the vulnerability, or failure on output that does.
\emph{Environment Setup Failure}: the agent could not bring the container to a runnable state, so no payload ran. This typically arises because an old version's libraries and build tools, once easy to install, are now hard to resolve.
\emph{Illegitimate Migration}: the agent passed by inventing a nonexistent API or parameter, or by editing the framework source, rather than exploiting the shipped version; the prompt forbids these (Section~\ref{sec:method-rq2}), so this category captures circumvented instructions.

Table~\ref{tab:rq3-failures} in Section~\ref{app:failure-distributions} reports the distribution of failures across these five categories for each model and harness. Because configurations with higher resolve rates fail on fewer instances, equal percentages correspond to different absolute counts.

To describe how the remaining failures differ across configurations, we correlate each model's resolve rate (from Section~\ref{sec:rq2}) with its share of failures in each category. 
Table~\ref{tab:rq3-corr} reports Spearman's $\rho$ and Kendall's $\tau_a$ per category and per harness, following the methodology of Appendix~\ref{app:failure-protocol}.
The directions agree across harnesses, but both evaluate the same models and instances. These are descriptive associations with benchmark resolve rate, not causal estimates of model capability.

\begin{table}[!htbp]
  \centering
  \captionsetup{font=small,skip=4pt}
  \caption{Rank correlation between a model's resolve rate and its share of failures in each category, computed separately for each harness across the nine models. We report Spearman's $\rho$ and Kendall's $\tau_a$, with tied ranks averaged.}
  \label{tab:rq3-corr}
  \footnotesize
  \setlength{\tabcolsep}{3pt}
  \renewcommand{\arraystretch}{1.0}
  \begin{tabular}{@{}p{0.39\columnwidth}cccc@{}}
    \toprule
    & \multicolumn{2}{c}{OpenHands} & \multicolumn{2}{c}{SWE-agent} \\
    \cmidrule(lr){2-3}\cmidrule(lr){4-5}
    \raggedright Failure category & $\rho$ & $\tau_a$ & $\rho$ & $\tau$ \\
    \midrule
    Missed API Migration            & $-0.93$ & $-0.83$ & $-0.87$ & $-0.72$ \\
Correct Strategy but Incomplete & $+0.93$ & $+0.83$ & $+0.93$ & $+0.83$ \\
Oracle Misinterpretation        & $-0.90$ & $-0.78$ & $-0.82$ & $-0.67$ \\
Environment Setup Failure       & $-0.03$ & $0.00$ & $-0.08$ & $-0.03$ \\
Illegitimate Migration          & $-0.41$ & $-0.25$ & $-0.37$ & $-0.14$ \\
    \bottomrule
  \end{tabular}
  \par\vspace{2pt}
  {\scriptsize\raggedright Negative values mean the category becomes a smaller share of failures as benchmark resolve rate increases; positive values mean it becomes a larger share. $\rho$~=~Spearman, $\tau_a$~=~Kendall.\par}
\end{table}

Higher resolve rates are associated with a smaller share of Missed API Migration failures ($\rho=-0.93$ for OpenHands and $-0.87$ for SWE-agent) and a larger share of Correct Strategy but Incomplete failures ($\rho=+0.93$ in both). Kimi-2.5 departs from this ordering in both harnesses. Because the analysis conditions on failure, a growing category share does not imply more failures in absolute terms. This suggests that completing and verifying a migration remains difficult even when an agent identifies the intended approach.
Illustrative failure trajectories are provided in Sections~\ref{app:failure-cases} and~\ref{app:dependency-case}.

Environment Setup Failure accounts for 15.5\%--19.3\% of failures across configurations and has little rank correlation with resolve rate ($\rho=-0.03$ for OpenHands and $-0.08$ for SWE-agent). This is a stable share of remaining failures, not a constant failure rate over all instances. It does not establish that environment repair is independent of model capability.

Oracle Misinterpretation becomes a smaller share of failures as resolve rate increases ($\rho=-0.90$ for OpenHands and $-0.82$ for SWE-agent). This suggests that self-assessment deserves attention when agents use their own verdict to stop, while the benchmark's independent checks remain necessary.

Illegitimate Migration accounts for 1.8\%--4.3\% of failures. These outputs violate task constraints, for example by changing the affected framework rather than adapting the PoC. The observed frequency describes rejected outputs; it does not establish whether the violations were intentional.

\subsubsection{Failure Analysis}
\label{app:failure-distributions}
Table~\ref{tab:rq3-failures} reports the distribution of failure root causes
across all agent--model configurations.

\begin{table}[!htbp]
  \centering
  \captionsetup{font=small,skip=4pt}
  \caption{Distribution of failure root causes. Parentheses after each model give its absolute number of failures out of 127.}
  \label{tab:rq3-failures}
  \footnotesize
  \setlength{\tabcolsep}{2pt}
  \renewcommand{\arraystretch}{1.0}
  \begin{tabularx}{0.88\columnwidth}{@{}llYYYYY@{}}
    \toprule
    Agent & Model & IM & OM & ES & MM & CI \\
    \midrule
    \multirow{9}{*}{\rotatebox{90}{OpenHands}}
      & GPT-OSS-120b (110) & 3.6 & 13.6 & 19.1 & 28.2 & 35.5 \\
      & Qwen-3.5 (96)      & 3.1 & 11.5 & 16.7 & 27.1 & 41.7 \\
      & Kimi-2.5 (94)      & 2.1 & 16.0 & 17.0 & 30.9 & 34.0 \\
      & GLM-5.2 (71)       & 2.8 &  9.9 & 16.9 & 23.9 & 46.5 \\
      & GPT-5.3 (70)       & 4.3 &  7.1 & 17.1 & 20.0 & 51.4 \\
      & Grok-4.6 (67)      & 3.0 &  9.0 & 16.4 & 20.9 & 50.7 \\
      & Gemini-3.1 (66)    & 3.0 &  7.6 & 16.7 & 16.7 & 56.1 \\
      & Sonnet-4.6 (63)    & 3.2 &  6.3 & 17.5 & 15.9 & 57.1 \\
      & Opus-4.6 (53)      & 1.9 &  5.7 & 17.0 & 15.1 & 60.4 \\
    \midrule
    \multirow{9}{*}{\rotatebox{90}{SWE-agent}}
      & GPT-OSS-120b (114) & 3.5 & 14.9 & 19.3 & 24.6 & 37.7 \\
      & Qwen-3.5 (97)      & 3.1 & 11.3 & 15.5 & 25.8 & 44.3 \\
      & Kimi-2.5 (94)      & 2.1 & 18.1 & 16.0 & 27.7 & 36.2 \\
      & GLM-5.2 (73)       & 2.7 &  9.6 & 16.4 & 24.7 & 46.6 \\
      & GPT-5.3 (72)       & 2.8 &  6.9 & 16.7 & 22.2 & 51.4 \\
      & Grok-4.6 (68)      & 2.9 &  8.8 & 16.2 & 22.1 & 50.0 \\
      & Gemini-3.1 (66)    & 3.0 &  7.6 & 16.7 & 15.2 & 57.6 \\
      & Sonnet-4.6 (64)    & 3.1 &  6.2 & 17.2 & 14.1 & 59.4 \\
      & Opus-4.6 (56)      & 1.8 &  7.1 & 16.1 & 14.3 & 60.7 \\
    \bottomrule
  \end{tabularx}
  \par\vspace{2pt}
  {\scriptsize\raggedright Rows sum to 100\% up to rounding; stronger models have fewer total failures, so equal percentages correspond to different absolute counts. IM~=~Illegitimate Migration, OM~=~Oracle Misinterpretation, ES~=~Environment Setup Failure, MM~=~Missed API Migration, CI~=~Correct Strategy but Incomplete.\par}
\end{table}

\subsubsection{Failure Case Studies}
\label{app:failure-cases}
CVE-2023-44389~\cite{CVE-2023-44389}, a stored cross-site scripting flaw in Zope's management interface, is a clear instance of the first failure type.
The reference PoC attacks a page that Zope~4 introduced, and that page does not exist in the unlisted target version 2.13.28.
The vulnerability is still reachable there, but through a different page: \texttt{manage\_main} writes an object title into the HTML \texttt{<title>} element using a template directive that does not escape its input, so a stored payload is returned unchanged and the oracle records a successful exploit.
The migration therefore comes down to finding the right page.
Kimi-2.5 on SWE-agent stored the payload without trouble and then went looking, but every page it tried was the wrong one: a redirect that returned no content, an outer frame that renders no title, and two pages that escape the title before displaying it.
Its first attempt was the near miss.
That redirect pointed at \texttt{manage\_main}, so the correct page was named in a header the agent's client was configured not to follow, and the agent discarded the lead after seeing an empty response.
It then switched to searching for the markup used by the reference version, found none, and concluded that the version was safe.
The reasoning went wrong in one step: the agent took ``the page from the reference PoC is absent'' to mean ``no unescaped title sink exists here.''
The first is true and the second is not, and it is that conflation, rather than a refusal to adapt, that Missed API Migration captures.
 
The second failure type is illustrated by CVE-2025-66471~\cite{CVE-2025-66471}, a decompression-bomb vulnerability (CWE-400) in urllib3, where Opus-4.6 on SWE-agent got the strategy right but could not finish.
The reference exploit relies on urllib3's streaming API, which does not exist in the target version 0.4.1.
Opus correctly recognized this and switched to the one-shot read path that 0.4.1 exposes, which is the substantive part of the migration.
It then failed on verification: to confirm a decompression bomb one must observe memory growth in the process that performs the decompression, which is the server, but Opus measured resident memory in the client process that only holds the HTTP response object.
Across its remaining iterations, it adjusted the bomb size, added a delay, and lowered its detection threshold, never recognizing that it was measuring the wrong process, and concluded that 0.4.1 was not vulnerable.
The exploitation strategy was correct end-to-end. Only the final measurement, in the wrong place, prevented the agent from confirming a vulnerability that was in fact present.

\subsubsection{Dependency Conflict Case Study}
\label{app:dependency-case}

The aiohttp 3.10.1 case (CVE-2024-42367~\cite{CVE-2024-42367}) illustrates the difficulty. Across five build attempts, dependency pins exposed further conflicts involving \texttt{yarl}, \texttt{multidict}, \texttt{frozenlist}, and build tools (Table~\ref{tab:rq3-depconflict}). The attempted configurations did not produce a runnable container, so the agent never tested the exploit. This trajectory motivates better support for historical environments, but does not prove that no compatible configuration exists.

\begin{table}[!htbp]
  \centering
  \setlength{\tabcolsep}{4pt}
  \renewcommand{\arraystretch}{1.0}
  \captionsetup{font=small,skip=4pt}
  \caption{The dependency conflict chain the agent faced on CVE-2024-42367 (target aiohttp 3.10.1). The attempted pins expose further conflicts; none of these attempts produces a runnable container.}
  \label{tab:rq3-depconflict}
  \footnotesize
  \begin{tabular}{@{}cp{2.6cm}p{3.9cm}@{}}
    \toprule
    Attempt & Pin applied & Conflict it exposed \\
    \midrule
    1 & none (initial build) & \texttt{yarl} resolves to 2.0.0, too new for aiohttp 3.10.1 \\
    2 & \texttt{yarl==1.9.4} & \texttt{multidict} resolves to 7.0.1, above aiohttp's ceiling \\
    3 & \texttt{multidict==6.0.5} & \texttt{frozenlist} 1.5.0 needs \texttt{setuptools>=68} \\
    4 & upgrade \texttt{setuptools} & \texttt{setuptools>=70} breaks \texttt{aiosignal} 1.3.1 \\
    5 & \texttt{setuptools==69.5}, \texttt{frozenlist==1.4.1} & \texttt{frozenlist} 1.4.1 needs \texttt{Cython}, which needs \texttt{setuptools<66} (circular) \\
    \bottomrule
  \end{tabular}
\end{table}

\find{\textbf{Answer to RQ3:} Higher resolve rates are associated with a smaller share of missed-API failures and a larger share of incomplete executions among the remaining failures, so the dominant failure mode shifts from locating the right entry point to completing a correct execution.}

\section{Discussion}
\subsection{Implications}
\label{sec:lessons}
The benchmark separates several requirements for migration: adapting an entry point, finding an alternative path, reconstructing an environment, and verifying exploitation. 
Attack-vector shifts have the lowest maximum resolve rate across categories. Among the remaining failures, higher-performing configurations more often identify an approach but fail to complete it. These results motivate evaluating intermediate execution and verification behavior alongside final success.

Environment failures also motivate better support for historical dependencies. Their stable share does not show that model improvements cannot help; testing that claim would require controlled interventions. Similarly, vulnerability-specific oracles and causal review serve different purposes: the oracle detects an observable effect, while review checks its relation to the disclosed flaw and the task constraints.

For metadata auditing, unchanged reference PoCs confirm only 22 of the 149 vulnerable versions identified here. Migration extends this evidence to another 127 versions. The benchmark evaluates adaptation on confirmed positive cases, so deploying it as an auditing method would additionally require evaluation on unaffected and inconclusive versions.

\noindent\textbf{Limitations.}
Our evaluation covers 127 validated migrations across 45 Python web CVEs, with one run per configuration and instance. It measures adaptation on confirmed vulnerable versions, rather than auditing accuracy on arbitrary versions. Comparisons depend on the supplied reference materials, available tools, and execution budget. Section~\ref{sec:threats} discusses construction bias, validation, baseline comparability, and generalization in detail.

\subsection{Limitations and Evaluation Scope}
\label{sec:threats}
\textbf{Internal validity.} Phase~1 and~2 rely on manual work, so an author could mismark an exploitable version. 
The dual independent attempts and the validation protocol of Section~\ref{sec:validation} make false accepts unlikely, but false rejects are not similarly controlled: they arise both when the two authors exhaust the 60-minute budget and when a candidate version's historical environment could not be built, leaving its status undetermined (the \textit{migratable failure}, \textit{timeout}, and \textit{unbuildable} segments of Table~\ref{tab:ledger}). Our 149 confirmed versions are therefore a \emph{lower bound} on under-reporting within the five-version window.
Because the benchmark contains only versions for which we produced a validated migration, it measures sensitivity on known-positive tasks and does not estimate false-positive rate, end-to-end audit specificity, or an ecosystem-wide prevalence rate.
Each tool is run once per instance, so run-to-run variance is unmeasured. Repeated runs are needed to assess the stability of aggregate and per-category differences. The same validators construct and evaluate instances; blinding to model identity reduces one source of bias but does not eliminate shared judgment errors. 
Finally, \vsw{} is one possible port of VulScope to Python web, not the definitive one, so its 8.7\% should be read as a result for this adaptation rather than a ceiling for trace-guided migration.

\textbf{External validity.} The benchmark covers 45 CVEs in Python web frameworks, and the category distribution, under-reporting rate, and absolute resolve rates may differ in other ecosystems and will shift as models improve.
The observed correlations use nine models per harness and failure shares that sum to one. They may reflect changes in which instances remain unresolved, as well as differences between models. Their persistence across ecosystems or future model generations remains to be tested.

\subsection{Ethical Considerations}
\label{sec:ethics}
All CVEs in \bench are publicly disclosed and have public fixes, and every migrated PoC targets a historical release and was executed only inside the containerized evaluation environments we built (Section~\ref{sec:validation}).
The benchmark does not discover new vulnerabilities: it establishes that already-known flaws also reach versions that the databases list as unaffected, which is precisely the information defenders need to correct affected-version metadata.

\section{Related Work}
\label{sec:related-work}

\subsection{PoC and Exploit Migration}
VulScope migrates C/C++ PoCs by aligning execution traces and mutating inputs~\cite{vulscope}; AEM transfers exploit strategies across Linux kernel versions using slab-layout conditions~\cite{jiang2023aem}. Patch-guided techniques use changes in vulnerable code to guide exploit reproduction~\cite{yang20231dfuzz,peng20191dvul}. Diffploit uses LLMs to migrate Java library exploits, evaluating 689 tasks from 102 CVEs with executable tests~\cite{chen2025diffploit}. \bench extends this evaluation to Python web software, where migration can require changes to both the PoC and its environment. Its validation combines vulnerability-specific execution checks with manual review of whether the disclosed flaw caused the observed effect.

\subsection{Cybersecurity Benchmarks for AI Agents}
\label{sec:rw-cyber}
 
Table~\ref{tab:bench-compare} in Section~\ref{app:comparison} compares \bench with related benchmarks. Cybench and NYU CTF Bench evaluate agents on capture-the-flag tasks~\cite{zhang2025cybench,shao2025nyuctfbenchscalable}. Benchmarks based on disclosed vulnerabilities evaluate reproduction or exploitation at a specified revision, e.g. ARVO~\cite{mei2026arvo} and CyberGym~\cite{ wang2026cybergym}, which also identify cases where reported fixes are incomplete. In the web setting, CVE-Bench evaluates 40 CVEs against eight attack goals~\cite{zhu2025cvebench}, and Liu et al.\ study reproduction of 80 web CVEs using vulnerability-specific oracles~\cite{liu2025webrepro}. 
Different from these existing benchmarks, \bench requires adapting a working PoC to an earlier, unlisted version, which tests how agents reason about cross-version code changes and yields evidence of  under-reported vulnerable versions.

\subsection{Coding Benchmarks for AI Agents}
\label{sec:rw-coding}
 
SWE-bench evaluates issue resolution using repository tests~\cite{jimenez2024swe}, while Multi-SWE-bench extends evaluation across languages~\cite{zan2025multiswe}. SWT-bench evaluates test generation~\cite{muendler2024swtbench}, and Terminal-Bench evaluates tasks in stateful terminal environments~\cite{merrill2026terminal}. Version-sensitive coding and migration benchmarks examine compatibility under changing dependencies~\cite{misra2026gitchameleon,islam2023pymigbench}. \bench shares their use of executable feedback, but requires the adapted program to preserve a vulnerability's exploitation behavior across versions. This includes cases where the reference entry point no longer reaches the vulnerable code.

\section{Conclusion and future work}
We introduce \bench, a benchmark of 127 validated cross-version PoC migrations across 45 Python web CVEs, to evaluate how agents adapt a working solution while preserving its target behavior. The best of 18 agent configurations resolves 58.3\% of instances. Attack-vector shifts remain difficult, and incomplete execution accounts for a larger share of the remaining failures of higher-performing configurations. The reference-package ablation further shows that supplied examples and environment information substantially support one evaluated configuration. Together, these findings motivate studying adaptation, execution, and verification as distinct requirements for successful agent behavior. As a practical outcome, benchmark construction also confirms 149 vulnerable prior versions absent from four affected-version databases. Future work will extend the benchmark across ecosystems, assess repeated-run stability, and include unaffected versions for metadata auditing.

\bibliographystyle{ACM-Reference-Format}
\bibliography{main}

\clearpage
\appendix

\section{Oracle Patterns}
\label{app:oracle-patterns}

Representative oracle patterns include the following.
\textbf{SQLi}: the HTTP response body contains a pre-planted canary row drawn from a controlled table.
\textbf{XSS}: the response body contains the injected script verbatim and is served with \texttt{Content-Type: text/html}.
\textbf{Path traversal}: a file at a declared path outside the intended root becomes readable through the vulnerable endpoint.
\textbf{RCE}: a file at a pre-declared path inside the container appears, created by an executed payload.
\textbf{ReDoS}: the end-to-end response latency for the crafted input exceeds a pre-registered threshold, while a benign control input stays below a second threshold.

\section{Reference PoC Construction Details}
\label{app:reference-details}

The two annotators have three or more years of active competitive Capture-the-Flag experience, and one has three years of industry experience as a security research engineer.
 
For each CVE, the annotators studied the advisory and its patch commit, and drew on any publicly available evidence they could find to reconstruct the vulnerability: vulnerability databases~\cite{osv,snyk}, public collections that link CVEs to community-posted PoC exploits, pre-built vulnerable-environment collections, and the disclosure platforms that NVD commonly references~\cite{huntr,exploitdb}.
They also used coding assistants to help with routine setup work, such as writing a basic Dockerfile or assembling a standard HTTP client.
 
To ensure reproducibility and support automated evaluation, the annotators manually record each PoC in a uniform JSON specification with four components. The first is the environment setup: the target software version, dependency packages, and Docker configuration required to implement the vulnerable application.
The second is the set of crafted artifacts: payloads, overridden configuration files, and auxiliary scripts.
The third is the execution instructions: the ordered steps to deploy the environment, deliver the exploit, and observe the outcome.
The fourth is the success oracle, the per-PoC condition that witnesses a successful exploit, defined and validated as described in Section~\ref{sec:validation}.

\section{Trace-Guided Baseline Adaptation}
\label{app:baseline-adaptation}

\textit{\vsw}
VulScope~\cite{vulscope} is the state-of-the-art trace-guided migration tool, designed for C/C++ user-space binaries, which relies on LLVM instrumentation, fuzzing, and crash-based triage. 
These mechanisms do not transfer to Python web applications: Python is interpreted rather than compiled, many framework calls enter opaque C extensions, and successful exploitation is observed through application-level effects rather than OS-level crashes. 
We therefore build \vsw, which keeps VulScope's core migration logic but replaces the low-level components specific to C/C++ binaries. 
It preserves VulScope's three defining properties: it aligns reference and target execution traces across versions, reasons about execution detours, and mutates \emph{only} PoC inputs while prioritizing candidates by their similarity to the reference trace. 
The detour signal is also adapted: 
in place of VulScope's unexpected-crash detour, 
\vsw uses an early-rejection detour, 
where a validator, guard, parser, or dispatcher rejects the request before it reaches the vulnerable sink. 
The full trace abstraction, cross-version function mapping, detour reasoning, and taint procedure are detailed in Appendix~\ref{app:vulscopeweb}.

\section{Run Budget Rationale}
\label{app:budget}

The budget limit follows the per-trajectory cap used by McFadyen et al.~\cite{mcfadyen2026inference} for TerminalBench~\cite{merrill2026terminal}, the closest published analogue to our setting in that it combines terminal access, a stateful containerized environment, and programmatic binary verification. 
That work reports that stateful, tool-using benchmarks continue to improve well beyond typical published budgets, so smaller caps risk mislabeling instances as non-migratable when the agent has merely exhausted its budget. 

\section{Failure Analysis Protocol Details}
\label{app:failure-protocol}

\textit{Failure pool.} The failure pool is the set of (instance, tool) pairs that the validators rejected during the output acceptance procedure (Section~\ref{sec:method-rq2}). 
We categorize only LLM-agent failures. The input-only baseline has a restricted action space, so its results are analyzed separately by migration category.

\textit{Coding protocol.} 
We assigned each failure a single root-cause category through open card sorting~\cite{spencer2009card}. 
The two validators who performed the output acceptance independently coded a random $20\%$ sample of failures using free-text descriptions, met to merge their descriptions into a draft codebook, refined it until the categories were stable and mutually exclusive, and applied the final codebook to the remaining failures, with disagreements resolved through discussion to consensus. 
The resulting five-category codebook is presented with its distribution in Section~\ref{sec:rq3}.

\textit{Correlation analysis.} 
To test whether the failure distribution shifts with model capability, we correlate each model's resolve rate with its share of failures in each category. 
We use rank correlation rather than a linear fit, both because we care about monotonic association and because each harness contributes only nine models, and we report Spearman's $\rho$ and Kendall's $\tau_a$.
Because each model appears once per harness and the two appearances are not independent, we compute the correlations separately for OpenHands and SWE-agent, which also lets the harnesses serve as mutual robustness checks: a pattern holding in both is more likely a property of model capability than an artifact of a single harness.

\section{\vsw Implementation Details}
\label{app:vulscopeweb}

\vsw represents an execution as an ordered tree of function invocations, following VulScope's trace abstraction. Each node corresponds to a Python function invocation keyed by its fully qualified name (\texttt{module.qualname}, and \texttt{module.Class.method} for methods), and inside each call, we record the ordered line numbers or bytecode offsets executed, with the callsite line recorded on each edge. We avoid file paths and line ranges as identities since line numbers shift across versions and destabilize alignment.
Two Python-specific cases require normalization: calls into C extensions, which cannot be traced at the Python level, are kept as labeled leaf nodes, and comprehensions and closures are merged into their parent function so equivalent executions still match across versions.

For cross-version function mapping, \vsw first maps functions with identical qualified names. 
For renamed or moved functions, it computes normalized AST similarity combined with Jaccard~\cite{jaccard} similarity over callers, callees, and constants, accepting a pair when the score exceeds $0.7$; when patch information is available, the buggy site is additionally mapped from the patch commit.
Unlike VulScope, \vsw does not implement the function split/merge transformation, which we treat as a baseline limitation: VulScope targets broader version gaps where such refactorings are common, whereas our benchmark targets only the five most recent versions before the earliest affected version, where direct, rename, and move mappings capture the intended operating regime.

After mapping, \textsc{VulScope-Web} aligns the reference and target trace trees and uses VulScope's seed-prioritization score
\[
  \mathrm{sim}(T_\mathit{ref}, T_\mathit{target}) =
  \frac{|\mathit{MATCHED}_\mathit{ref}|}
       {|V_\mathit{ref}| - |\mathit{NOTCORRECTABLE}_\mathit{ref}|},
\]
where $\mathit{MATCHED}_\mathit{ref}$ is the set of reference nodes aligned to the target trace, $V_\mathit{ref}$ is all reference nodes, and $\mathit{NOTCORRECTABLE}_\mathit{ref}$ contains reference-side detours judged impossible to correct on the target version. 
Although the implementation uses Atheris~\cite{atheris} for byte-level mutation operators, queue ordering follows this trace-similarity objective rather than Atheris's default coverage objective, so the baseline is a Python implementation of VulScope's trace-guided input-adjustment loop, not a generic coverage-guided fuzzer.

\textsc{VulScope-Web} also ports VulScope's detour reasoning, detecting missed calls (a reference-trace function absent from the target trace) and unintended calls (a target-trace function not aligned with the reference).
For each correctable detour, it performs backward dominator analysis over the parent frame's executed line or bytecode sequence to identify the gating branch condition or argument, then localizes that critical variable to input fields using a lightweight hybrid taint procedure: static taint propagates from request fields to branch conditions and patched-function arguments along the aligned call chain, and when static taint cannot connect a field to the critical variable, \textsc{VulScope-Web} falls back to differential field probing, perturbing one input field at a time and observing whether the detour or the trace-similarity score changes.

\end{document}